\documentclass[aps,prl,twocolumn,showpacs,amsmath,amssymb,floatfix,superscriptaddress]{revtex4-1}
\usepackage{graphicx}
\usepackage{dcolumn}
\usepackage{bm}
\usepackage{color}
\usepackage{graphicx}
\usepackage{epstopdf}
\usepackage{epsfig}
\usepackage[colorlinks=true, citecolor=blue, anchorcolor=green]{hyperref}

\include{graphic}

\makeatletter
\begin{document}
\preprint{APS/123-QED}

\title{Magnonic time crystal with lasing coherence}

\author{Jiantao~Song}
\affiliation{School of Physical Science and Technology, ShanghaiTech University, Shanghai 201210, China}
\affiliation{State Key Laboratory of Infrared Physics, Shanghai Institute of Technical Physics, Chinese Academy of Sciences, Shanghai 200083, China}

\author{C.K.~Wei}
\affiliation{State Key Laboratory of Infrared Physics, Shanghai Institute of Technical Physics, Chinese Academy of Sciences, Shanghai 200083, China}

\author{Shuhuan~Ma}
\affiliation{School of Physical Science and Technology, ShanghaiTech University, Shanghai 201210, China}
\affiliation{State Key Laboratory of Infrared Physics, Shanghai Institute of Technical Physics, Chinese Academy of Sciences, Shanghai 200083, China}

\author{Yiming~Pan}
\email{yiming.pan@shanghaitech.edu.cn}
\affiliation{School of Physical Science and Technology, ShanghaiTech University, Shanghai 201210, China}

\author{Zhijian~Chen}
\affiliation{School of Physical Science and Technology, ShanghaiTech University, Shanghai 201210, China}
\affiliation{State Key Laboratory of Infrared Physics, Shanghai Institute of Technical Physics, Chinese Academy of Sciences, Shanghai 200083, China}

\author{Li~Yu}
\affiliation{State Key Laboratory of Infrared Physics, Shanghai Institute of Technical Physics, Chinese Academy of Sciences, Shanghai 200083, China}

\author{Wei~Lu}
\email{luwei@shanghaitech.edu.cn}
\affiliation{State Key Laboratory of Infrared Physics, Shanghai Institute of Technical Physics, Chinese Academy of Sciences, Shanghai 200083, China}
\affiliation{School of Physical Science and Technology, ShanghaiTech University, Shanghai 201210, China}

\author{Bimu~Yao}
\email{yaobimu@mail.sitp.ac.cn}
\affiliation{State Key Laboratory of Infrared Physics, Shanghai Institute of Technical Physics, Chinese Academy of Sciences, Shanghai 200083, China}
\affiliation{School of Physical Science and Technology, ShanghaiTech University, Shanghai 201210, China}

\begin{abstract}

Discrete time crystals (DTC) can respond at an integer multiple of the drive period. Internal order-parameter measurements have advanced studies of this novel non-equilibrium phase of matter; however, turning this order into highly coherent, externally accessible radiation remains challenging: pump-driven nonlinear processes must supply sufficient gain to compensate dissipation while preserving subharmonic stability. Here we realize a room-temperature magnonic time crystal (MTC) with radiation of lasing coherence using a pump-induced magnon mode in yttrium iron garnet. A microwave beat drive produces robust period doubling over a finite parameter range, with spectral signatures of higher-order MTC at stronger driving. The emitted field exhibits linewidth narrowing to $260.0~\mathrm{Hz}$ and a first-order correlation decay time of $995~\mu\mathrm{s}$.  Our model links this response to three-magnon confluence, yields a stability region consistent with experiment and predicts a possible transition to chaos. We further map the complex temporal dynamics into frequency-resolved nonlinear features for reservoir computing. These results bridge time-crystalline order and nonlinear information processing through intrinsic magnon–magnon interactions and lasing coherence.

\end{abstract}

\maketitle

\textit{Introduction}.---Spatial crystals organize matter periodically in space\cite{LIB-CHAIKIN1995-PrinciplesCondensedMatterPhysics}; time crystals extend this ordering into time \cite{TIME-WILCZEK12,LIB-WILCZEK2013-SuperfluiditySpaceTimeTranslation,LIB-KISELEV2024-LightControlledTerahertzPlasmonic}. In periodically driven systems, DTC spontaneously break discrete time-translation symmetry\cite{DTC-VANDERPOL27,LIB-SACHA2018-TimeCrystalsReview}, producing subharmonic oscillations\cite{DTC-VANDERPOL27} robust against weak perturbations\cite{DTC-ZHANG17}. DTC research spans quantum simulation \cite{LIB-RANDALL2021-ManyBodyLocalizedDiscrete}, non-equilibrium spin dynamics\cite{DTC-BEATREZ23,LIB-AUTTI2018-ObservationTimeQuasicrystalTransition,MAG-KREIL19,MAG-TRAEGER21}, and nonlinear and topological photonics\cite{DTC-TAHERI22B,LIB-WENG2026-DiscreteTimeCrystalsActively,DTC-PAN20,DTC-WANG22}. Recent advances include two-dimensional discrete time-crystalline order\cite{LIB-SWITZER2026-RealizationTwoDimensionalDiscrete}, temporal quasicrystallinity\cite{LIB-HE2025-ExperimentalRealizationDiscreteTime,LIB-ZHU2026-ObservationDiscreteTimeQuasicrystal} and topological time-crystalline order\cite{LIB-WILL2025-ProbingNonEquilibriumTopological}. Yet achieving highly coherent radiation from DTC in open electromagnetic environments remains challenging, constraining prospects for the implementation of time crystals in communicating with other devices in an on-chip architecture and integration of radiation for information processing\cite{LIB-VANDOORNE2014-ExperimentalDemonstrationReservoirComputing}. A central microscopic question is how a driven dissipative system accumulates the gain from the competition between pump, loss and nonlinearity. 

A related but distinct approach is found in photonic time crystals\cite{PTC-LYUBAROV22,LIB-ASGARI2024-TheoryApplicationsPhotonicTimeCrystals}, where periodic modulation creates momentum bandgaps that pump energy into electromagnetic modes\cite{DTC-KISELEV25}. This shows that temporal modulation can provide the gain needed for sustained radiation\cite{LIB-FEINBERG2025-PlasmonicTimeCrystals}. However, realizing this in conventional optical systems faces limitations\cite{LIB-GUO2026-PlasmonicMetamaterialTimeCrystal,LIB-WANG2025-ExpandingMomentumBandgapsPhotonic,LIB-SAHA2023-PhotonicTimeCrystalsMaterialsPerspective}. The nonlinearity in materials is usually weak\cite{LIB-WANG2025-ExpandingMomentumBandgapsPhotonic}, which limits the fast refractive-index modulation and thereby the gain accumulation rate\cite{LIB-HAYRAN2025-ResonantTonePhotonicTime,LIB-KHURGIN2021-FastSlowNonlinearitiesENZ,LIB-HAYRAN2022-DispersionEnergyConstraints}. Directly increasing external drive intensity could strengthen modulation but the materials suffer from severe loss and heating\cite{LIB-HAYRAN2022-DispersionEnergyConstraints}, while incorporating the external cavities to lower drive requirements comes with narrower operating bandwidth, limited rapid modulation speeds and tighter frequency-matching tolerances. This highlights the need for an alternative route—one that uses intrinsic, low-loss nonlinear interactions to accumulate energy into a coherent output without destroying the temporal order.

Here we address this challenge using a pump-induced magnon mode (PIM)\cite{LIB-RAO2023-UnveilingPumpInducedMagnon} in yttrium iron garnet (YIG) to realize a room-temperature magnonic time crystal with lasing microwave radiation. The YIG host provides a low-damping medium supporting highly coherent magnon states\cite{LIB-PIRRO2021-AdvancesCoherentMagnonics,MAG-KOSTER26}. Within this medium, the PIM introduces a coupling that is transiently tuned by the pump, which has previously enabled low-threshold nonlinear frequency conversion\cite{PIM-WANG24} and time-varying strong coupling\cite{PIM-RAO25}. The underlying three-magnon process transfers pump energy to secondary modes\cite{MAG-QU25,MAG-ARFINI25,MAG-QU23}, whose cross-branch confluence modify the temporal response and provide an effective gain.  Under a beating microwave drive, our system develops robust period doubling relative to the pump beat over a finite parameter range, followed by higher-order period multiplication at stronger driving. The emitted field exhibits linewidth narrowing\cite{LASER-SCHAWLOW58} to a few hundred hertz and correlations persisting for hundreds of microseconds, accompanied by a ring distribution in its IQ histogram\cite{LASER-RISKEN65,MASER-LIU15-SCIENCE,LIB-YAO2023-CoherentMicrowaveEmissionGain}. A nonlinear scattering model links the central radiation to three-magnon confluence and attributes the bounded period-doubled regime to competition between scattering channels. We further use the frequency-resolved output as  a high-dimensional temporal reservoir\cite{RC-JAEGER01,RC-MAASS02}, unfolding its complex periodic dynamics in frequency space and connecting temporal ordering with coherent microwave generation and readout.

\begin{figure*} [htpb]
\begin{center}
\epsfig{file=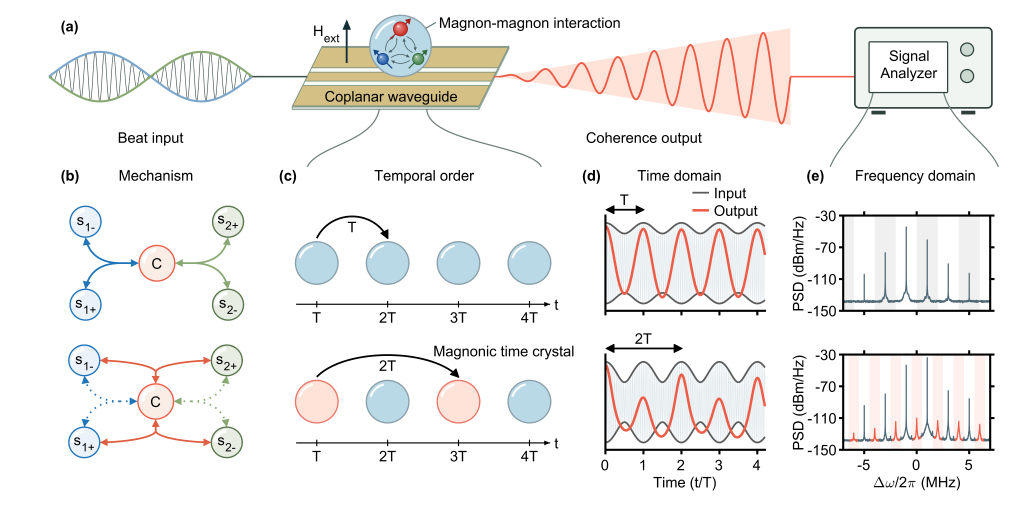,width=17.5cm} \caption{\textbf{Three-magnon interactions generate a period-doubled response.} \textbf{a,} Conceptual overview of a YIG sphere coupled to a coplanar waveguide (CPW), driven by two phase-coherent microwave components. Their beat note provides periodic driving, while nonlinear magnon interactions generate the coherent output recorded by a signal analyser.. 
\textbf{b,} Within-branch three-magnon processes (top) and additional cross-branch pathways (bottom, red), involving the high-frequency mode $C$ and secondary modes $s_{j\pm}$. \textbf{c,} Schematic recurrence of the $T$ and $2T$ responses, respectively. \textbf{d,} Reconstructed input (grey) and output (red) waveforms, with all spectral phases set to zero.
\textbf{e,} Measured radiation PSD at $f_1=2.611~\mathrm{GHz}$, $f_2=2.613~\mathrm{GHz}$ and $P_1=-3~\mathrm{dBm}$. The top and bottom spectra correspond to $P_2=-18$ and $8~\mathrm{dBm}$ and have comb spacings of $f_{\mathrm b}=2~\mathrm{MHz}$ and $f_{\mathrm b}/2=1~\mathrm{MHz}$, respectively. Frequency offsets are measured from $\bar f=2.612~\mathrm{GHz}$.}
\label{schematic}
\end{center}
\end{figure*}

\textit{Results}.---To realize a magnonic time crystal, we apply a microwave beat drive to a yttrium iron garnet (YIG) sphere placed on the centre conductor of a coplanar waveguide (CPW) at room temperature (Fig.~\ref{schematic}(a)). The drive comprises two phase-coherent frequency components at $f_1$ and $f_2$, with respective powers $P_1$ and $P_2$. Their separation $f_{\mathrm b}=f_2-f_1$ defines the beat period $T=1/f_{\mathrm b}$. The high-frequency magnon mode $C$ has frequency $f_c$, controlled by the bias magnetic field. We denote the mean drive frequency by $\bar f=(f_1+f_2)/2$. For the fixed-frequency measurements presented below, the two
drive components are placed symmetrically about the magnon mode, such that $\bar f=f_c$. The microwave output is collected through the CPW, and its power spectrum is recorded using a signal analyser (SA). Details of the microwave circuit are provided in the Supplementary Information.

The three-magnon processes in Fig.~\ref{schematic}(b) provide a microscopic picture of the temporal ordering illustrated schematically in Fig.~\ref{schematic}(c). In the upper panel of Fig.~\ref{schematic}(b), three-magnon splitting of the driven high-frequency mode $C$ populates secondary pairs $s_{j+}$ and $s_{j-}$ near $f_j/2$, where $j=1,2$ labels the two drive branches. Confluence within each branch returns energy to $C$ at the corresponding drive frequency. The associated response repeats every drive period $T$, as depicted in the upper panel of Fig. 1c. In the lower panel of Fig.~\ref{schematic}(b) the additional pathways highlighted in red connect the two branches: $s_{1+}$ combines with $s_{2-}$, and $s_{1-}$ with $s_{2+}$, feeding $C$ at the magnon-mode frequency $f_c = \bar f$. These pathways channel pump energy through the secondary magnons into a central frequency component that is not directly driven. In our model, feedback between this component and the secondary modes provides effective gain that can overcome damping and sustain the central radiation. Its frequency offset of $f_{\mathrm b}/2$ from either drive component supports a response that alternates between successive drive cycles and repeats only after $2T$ (Fig. ~\ref{schematic}(c), lower panel), linking microscopic gain to the period-doubled temporal order of the magnonic time crystal.

We characterize the temporal response through input and output envelopes reconstructed from measured microwave spectra, assigning zero phase to all spectral components (Fig.~\ref{schematic}(d)). The microwave beat drive comprises two phase-coherent frequency components at $f_1=2.611~\mathrm{GHz}$ and $f_2=2.613~\mathrm{GHz}$, whose separation sets the beat frequency $f_{\mathrm b}=2~\mathrm{MHz}$ and period $T=0.5 \mu\mathrm{s}$. With their power $P_1$ fixed at $-3~\mathrm{dBm}$, we compare two settings of $P_2$. At $P_2=-18 ~\mathrm{dBm}$, both the drive envelope (grey) and the output envelope (red) repeat every $T$ (upper panel). Increasing $P_2$ to $+8~\mathrm{dBm}$ produces an output envelope that alternates between successive drive cycles and repeats only after $2T$, while the drive remains $T$-periodic (lower panel). This alternation visualizes the period-doubled temporal order in the microwave output.

\begin{figure*} [htpb]
\begin{center}
\epsfig{file=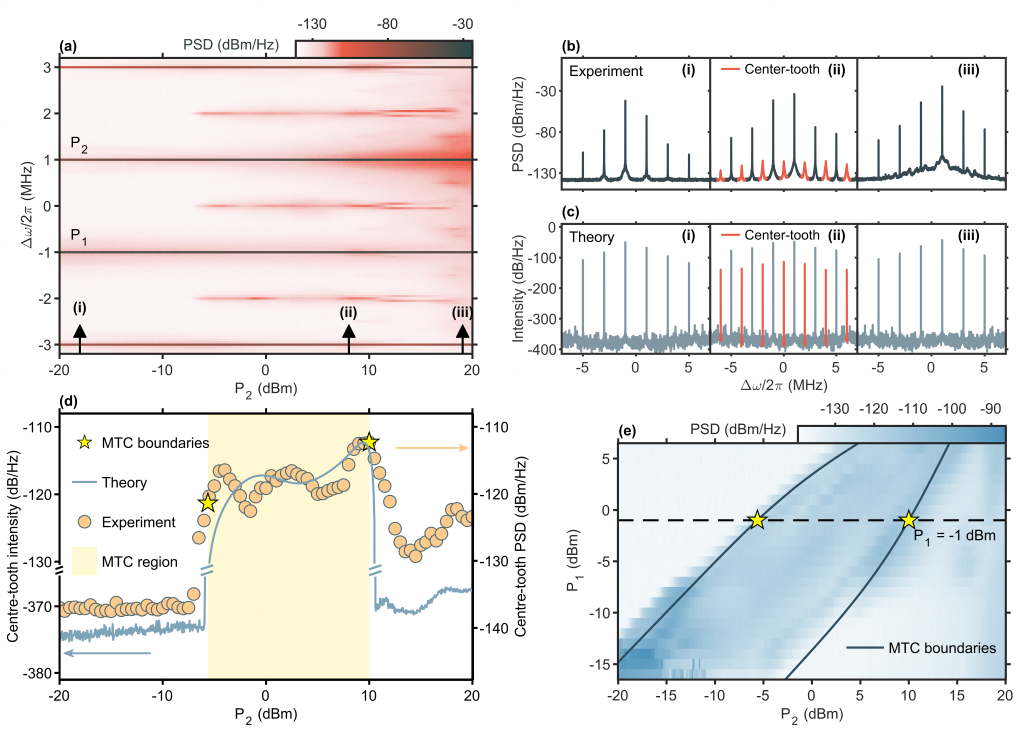,width=17.5cm}  
\caption{\textbf{Analytical stability boundaries delimit a bounded subharmonic regime.}
\textbf{a,} Experimental radiation spectrum versus pump-2 power $P_2$ at fixed $P_1=-1~\mathrm{dBm}$ and beat frequency $f_{\mathrm b}=2~\mathrm{MHz}$. Frequency offsets are measured from the mean pump frequency $\bar f$. Arrows indicate the three representative power settings.
\textbf{b,} Experimental spectra at $P_2=-18$, $8$ and $19~\mathrm{dBm}$, from left to right, showing the $1T$ comb, the interleaved $2T$ comb and high-power suppression of the intermediate lines. Highlighted teeth in (ii) identify the additional subharmonic components.
\textbf{c,} Calculated spectra at the same displayed pump-power coordinates as in \textbf{b}. 
\textbf{d,} Measured PSD at the centre frequency (symbols, right axis) and calculated PSD at the exact zero-offset Fourier bin (line, left axis). Yellow stars mark the approximate boundaries obtained from the local half-frequency perturbation criterion, and shading indicates the interval between them.
\textbf{e,} Centre-tooth PSD in the $(P_1,P_2)$ plane. Solid curves show the local stability estimates using the displayed power coordinates. The dashed line at $P_1=-1~\mathrm{dBm}$ identifies the scan in \textbf{d}; yellow stars locate the two boundary points marked in \textbf{d}.}
\label{TC}
\end{center}
\end{figure*}

The measured spectra in Fig.~\ref{schematic}(e) connect the period-doubled response to the emergence of a central radiation component. In the upper panel, the comb spacing equals the beat frequency $f_{\mathrm b}=2~\mathrm{MHz}$, corresponding to the $T$-periodic response. In the lower panel, a new peak appears at the undriven centre frequency $\bar f=2.612~\mathrm{GHz}$. We term the associated output the central subharmonic radiation, reflecting its frequency offset of $f_{\mathrm b}/2$ from either drive component. This peak emerges together with additional lines interleaved between the original comb teeth, reducing the spacing to $f_{\mathrm b}/2=1~\mathrm{MHz}$ and marking the transition to a $2T$ response. Its appearance is consistent with the cross-branch confluence pathway in Fig.~\ref{schematic}(b), linking the proposed microscopic gain mechanism to an externally accessible signature of magnonic time-crystalline order. The robustness and coherence of this central radiation are examined below.

We next examine the robustness of the period-doubled response and the parameter boundaries within which it persists. Keeping the pump and magnon-mode frequency settings of Fig.~\ref{schematic}, we fix $P_1=-1~\mathrm{dBm}$ and record the radiation spectrum while sweeping $P_2$ (Fig.~\ref{TC}(a)). At low $P_2$, the spectral map is dominated by dark stripes separated by $f_{\mathrm b}=1/T$, characteristic of the $T$-periodic response. As $P_2$ crosses an onset threshold, additional stripes, including the central subharmonic radiation, emerge abruptly between the original ones, reducing the spacing to $f_{\mathrm b}/2=1/(2T)$. This period-doubled regime persists over an intermediate power interval. Further increasing $P_2$ suppresses the intermediate stripes, restoring the original $1/T$ spacing. Three representative operating points are marked (i)–(iii) in Fig.~\ref{TC}(a), and their corresponding spectra are shown in Fig.~\ref{TC}(b). These spectra illustrate the initial $T$ response at $P_2=-18~\mathrm{dBm}$, the interleaved $2T$ comb at $P_2=8~\mathrm{dBm}$ and the recovery of the original comb spacing at $P_2=19~\mathrm{dBm}$. We therefore model the driven magnon dynamics to identify the microscopic processes governing both the onset and the high-power suppression of the central subharmonic radiation.

Starting from the effective-mode description of three-magnon scattering, we retain the driven high-frequency mode $\hat C$ and two half-frequency mode pairs $\hat s_{j\mu}$, where $j=1,2$ labels the pump branch and $\mu=\pm$ labels its momentum partners. We include confluence between branches through a sum over allowed partner pairs. The effective Hamiltonian is

\[
\begin{aligned}
\hat H&=\hat H_0+\hat H_d+\hat H_{3,\mathrm{reg}}+\hat H_{3,\times},\\
\frac{\hat H_0}{\hbar}&=\omega_c\hat C^\dagger\hat C+
\sum_{j=1}^2\sum_{\mu=\pm}\omega_{j\mu}\hat s_{j\mu}^\dagger\hat s_{j\mu},\\
\frac{\hat H_d}{\hbar}&=\frac{i}{2}\sum_{j=1}^2
\left[E_je^{-i\omega_jt}\hat C^\dagger-\mathrm{H.c.}\right],\\
\frac{\hat H_{3,\mathrm{reg}}}{\hbar}&=\sum_{j=1}^2
\left[\zeta_j\hat C^\dagger\hat s_{j+}\hat s_{j-}+\mathrm{H.c.}\right],\\
\frac{\hat H_{3,\times}}{\hbar}&=\sum_{(\mu,\nu)\in\mathcal P_\times}
\left[\zeta^\times_{\mu\nu}\hat C^\dagger\hat s_{1\mu}\hat s_{2\nu}
+\mathrm{H.c.}\right].
\end{aligned}
\label{eq:five_mode_H}\tag{1}
\]

The within-branch terms describe splitting into half-frequency partners and their reverse confluence. The set $\mathcal P_\times$ contains the allowed cross-branch pairs, whose coupling coefficients include mode overlap. Their products add coherently, so both amplitudes and phases enter the central source. For the numerical model, we make a further collective-amplitude reduction to $C,S_1,S_2$ within a minimal truncation with fixed symmetry-related partner profiles. Partner normalization and selected channel weights are absorbed into effective couplings, with $\zeta_\times$ denoting the effective coupling coefficient of the cross-branch three-magnon process. Adding linear and occupation-dependent losses gives

\[
\begin{aligned}
\dot C={}&-(\kappa_c+i\delta_c)C
-i(\zeta_1S_1^2+\zeta_2S_2^2+\zeta_\times S_1S_2)\\
&+F_1e^{+i\Omega t/2}+F_2e^{-i\Omega t/2},\qquad (2\mathrm{a})\\
\dot S_1={}&-(\gamma_1+q|S_1|^2+i\delta_{s1})S_1\\
&-iC(2\zeta_1^*S_1^*+\zeta_\times^*S_2^*),\qquad (2\mathrm{b})\\
\dot S_2={}&-(\gamma_2+q|S_2|^2+i\delta_{s2})S_2\\
&-iC(2\zeta_2^*S_2^*+\zeta_\times^*S_1^*).\qquad (2\mathrm{c})
\end{aligned}
\label{eq:meanfield_model}
\]

Here $C$ and $S_j$ are dynamical complex amplitudes of the high-frequency field and the two half-frequency fields, each of which can contain multiple frequency components. Frequency offsets are referenced to $\bar f$ and $\bar f/2$, respectively. All rates use angular-frequency units, $q=2\eta^2$, $\kappa_c$ and $\gamma_j$ are amplitude-decay rates, $\delta_c$ and $\delta_{sj}$ are detunings, and $F_j$ are calibrated drives. The two pumps lie on opposite sides of their mean frequency $\bar f$, separated by $f_{\mathrm b}$. The corresponding two half-frequency positions are $\bar f/2-f_{\mathrm b}/4$ and $\bar f/2+f_{\mathrm b}/4$, which sum to $\bar f$. When components of $S_1$ and $S_2$ occupy these complementary positions, their opposite time factors $e^{\pm i\Omega t/4}$ cancel in the product, where $\Omega=2\pi f_{\mathrm b}$. Cross-branch confluence then injects energy into the central frequency $\bar f$. The amplitudes and phases of the participating components determine how the allowed contributions add, while central-mode loss limits the resulting response.

Integrating Eqs.~(2a)--(2c) and Fourier-transforming the steady-state $C(t)$ reproduces the change in comb spacing from $1/T$ to $1/(2T)$ and back to $1/T$ (Fig.~\ref{TC}(c)). The measured and calculated PSDs at $f=\bar f$ both show a finite response interval at $P_1=-1~\mathrm{dBm}$ (Fig.~\ref{TC}(d)). For the principal central-confluence channel at $f_{\mathrm b}=2~\mathrm{MHz}$, the two half-frequency fields must contain components at opposite offsets of $-0.5$ and $+0.5~\mathrm{MHz}$ relative to $\bar f/2$. The pump-driven half-frequency responses compete as $P_2$ increases. In the intermediate interval, the lower-frequency component of $S_1$ coexists with the upper-frequency component of $S_2$. At higher power, the lower-frequency component of $S_1$ is suppressed and $S_1$ instead becomes dominated by the same upper-frequency response as $S_2$, suppressing the principal confluence channel that feeds the centre. The finite interval thus reflects competition that permits the complementary frequency components to coexist.

We expand the high-frequency field into its frequency components as
$C(t)=C_1e^{i\Omega t/2}+C_2e^{-i\Omega t/2}+C_0+\cdots$, where $C_1$ and $C_2$ are the pump-harmonic amplitudes including loading by the half-frequency response, $C_0$ is the central component at $\bar f$, and the ellipsis denotes other harmonics. On approximate reference backgrounds dominated by $S_1$ or $S_2$, we test the complementary $S_2$ or $S_1$ component, respectively. Linearizing Eqs.~(2a)--(2c) and eliminating auxiliary harmonics at zero envelope exponent gives the effective decay rates $\tilde\gamma_j$ and detunings $\tilde\delta_{sj}$, including feedback corrections. Combining the linear equations for each perturbation and its conjugate and taking the perturbation proportional to $e^{\lambda t}$ gives the largest real growth exponents within this local approximation

\[
\begin{aligned}
\lambda_j&=-\tilde\gamma_j+
\operatorname{Re}\sqrt{4|\zeta_j C_j|^2-\tilde\delta_{sj}^{\,2}}\
\qquad j=1,2.
\end{aligned}
\label{eq:weak_branch_response}\tag{3}
\]

\begin{figure*} [htpb]
\begin{center}
\epsfig{file=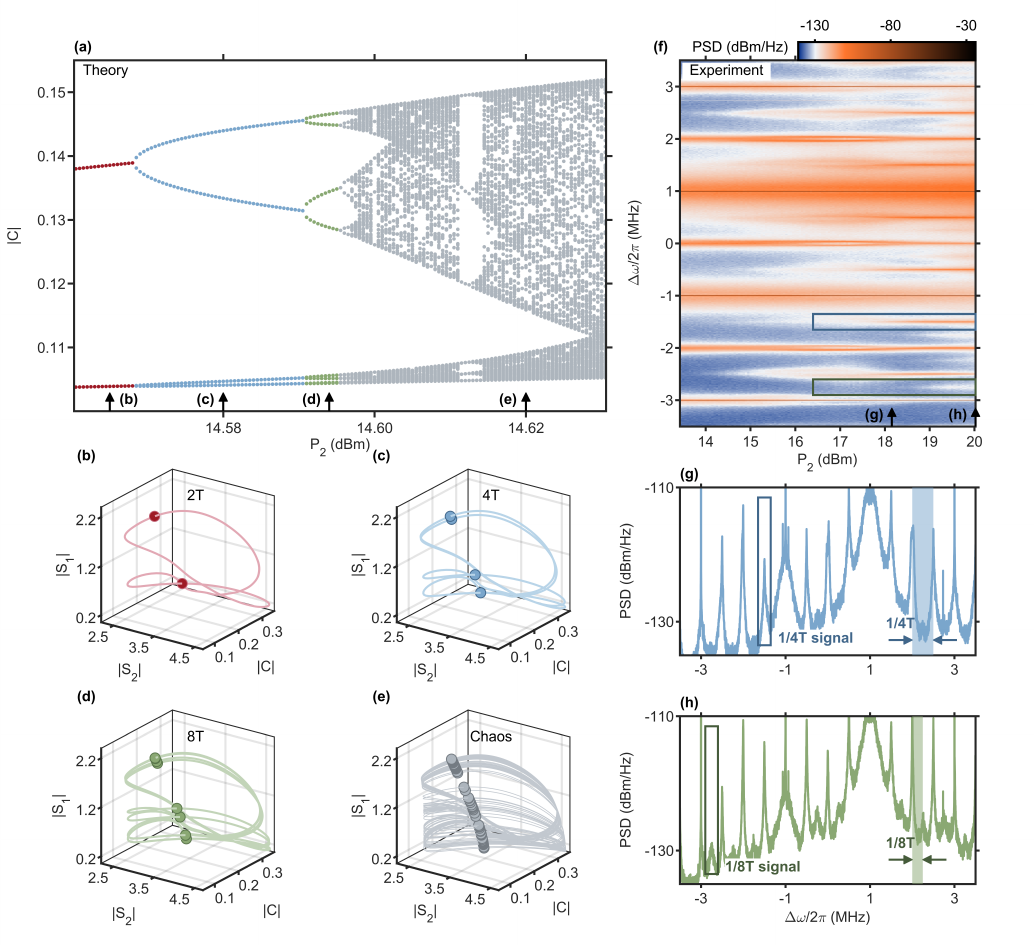,width=17.5cm}  \caption{\textbf{Higher-order spectral subdivision and period-multiplied model trajectories.}
\textbf{a,} Calculated bifurcation diagram of the central-mode amplitude $|C|$, sampled once per beat period at one fixed beat phase, with $P_1=6~\mathrm{dBm}$ and $f_{\mathrm b}=2~\mathrm{MHz}$. Red, blue and green points denote $2T$, $4T$ and $8T$ amplitude recurrence, respectively; grey points indicate other recurrence classes or aperiodic dynamics. Arrows identify the operating points in \textbf{b--e}.
\textbf{b--e,} Continuous post-transient trajectories in $(|C|,|S_1|,|S_2|)$ at $P_2=14.565$, $14.580$, $14.594$ and $14.620~\mathrm{dBm}$, respectively. Filled circles denote samples taken at the same beat phase and separated by $T=0.5~\mu\mathrm{s}$. The periodic cases contain two, four and eight distinct amplitude points; the chaotic case has an extended sampled set. Numerical Lyapunov analysis supports chaos at the last working point.
\textbf{f,} Measured radiation PSD versus $P_2$ at $P_1=6~\mathrm{dBm}$ and $f_{\mathrm b}=2~\mathrm{MHz}$. Blue and green frames highlight additional components, and arrows identify the spectra in \textbf{g,h}.
\textbf{g,h,} Measured spectra at $P_2=18.3$ and $20~\mathrm{dBm}$, respectively, resolving components separated by $0.5~\mathrm{MHz}=f_{\mathrm b}/4$ and $0.25~\mathrm{MHz}=f_{\mathrm b}/8$. These provide spectral signatures of $4T$ and $8T$ responses. Model and experimental operating powers are compared qualitatively. Frequency offsets are measured from $\bar f$.}
\label{Mutiple}
\end{center}
\end{figure*}

A negative $\lambda_j$ means that the corresponding half-frequency perturbation decays, a positive value means that it grows, and $\lambda_j=0$ gives a local boundary. In Eq.~(3), $\lambda_2$ is evaluated on the $S_1$ reference background and $\lambda_1$ on the $S_2$ reference background; the amplitudes and effective coefficients are evaluated on the respective backgrounds.

At fixed $P_1$, the change of $\lambda_2$ from negative to positive gives the lower boundary, while the high-power change of $\lambda_1$ from positive to negative gives the upper boundary. The zero-growth conditions of these analytical expressions define the two boundary curves as functions of $P_1$ in Fig.~\ref{TC}(e). The dashed line and yellow stars in Fig.~\ref{TC}(e) locate the scan in Fig.~\ref{TC}(d) and its boundary points in the parameter plane. This analytical criterion approximately estimates the coexistence interval of the complementary frequency components, while the central spectrum follows from the full nonlinear solution. Details of the parameter fitting, analytical formulas and derivations are provided in the Supplementary Information.

To explore how temporal order evolves under stronger beat driving, we extend the numerical analysis of the driven–dissipative magnon model introduced above using Eqs.~(2a)--(2c). We fix $P_1=6~\mathrm{dBm}$ and $f_{\mathrm b}=2~\mathrm{MHz}$ and numerically integrate the equations of motion as $P_2$ increases. After discarding transients, we sample the central-mode amplitude $|C|$ once every beat period $T$, at a fixed drive phase, to construct the bifurcation diagram in Fig.~\ref{Mutiple}(a). The two branches of the $2T$ response (red) successively split into four (blue) and eight (green), indicating a period-doubling cascade from $2T$ to $4T$ and $8T$. At higher powers, dense bands of stroboscopic amplitudes emerge (grey), suggesting a further transition towards aperiodic dynamics and chaos.

The trajectories at the four operating points marked in Fig.~\ref{Mutiple}(a) resolve these dynamical regimes in the $(|C|,|S_1|,|S_2|)$ amplitude space (Fig.~\ref{Mutiple}(b–e)). Curves show the continuous evolution after transients, while circles denote samples taken once every beat period. At $P_2=14.565$, $14.580$ and $14.594~\mathrm{dBm}$, the trajectories cycle through two, four and eight distinct stroboscopic points, respectively (Fig.~\ref{Mutiple}(b–d)). The mode amplitudes thus recur after $2T$, $4T$ and $8T$, with no shorter recurrence period. At $P_2=14.620~\mathrm{dBm}$, this discrete recurrence is lost: the trajectory no longer closes, and the stroboscopic samples spread over an extended set of amplitudes (Fig.~\ref{Mutiple}(e)). The largest Lyapunov exponent is positive and remains so upon halving the integration time step, providing theoretical evidence for chaotic magnon dynamics following the period-doubling sequence.

Experiments reveal corresponding signatures of higher-order period multiplication in the emitted microwave field. With $P_1=6~\mathrm{dBm}$ and $f_{\mathrm b}=2~\mathrm{MHz}$, we measure the radiation spectrum while sweeping $P_2$ (Fig.~\ref{Mutiple}(f)). As the drive power increases, additional lines emerge between the existing comb teeth, further subdividing the spectrum. The additional components highlighted in blue and green frames are resolved in the representative spectra shown in Fig.~\ref{Mutiple}(g,h). At $P_2=18.3~\mathrm{dBm}$, the spectrum exhibits a line spacing of $0.5~\mathrm{MHz}=f_{\mathrm b}/4$ (Fig.~\ref{Mutiple}(g)). Increasing $P_2$ to $20~\mathrm{dBm}$ introduces finer spectral structure with a spacing of $0.25~\mathrm{MHz}=f_{\mathrm b}/8$ (Fig.~\ref{Mutiple}(h)). These spacings provide spectral signatures of $4T$ and $8T$ responses and agree qualitatively with the higher-order periodic states predicted by the model. The experimentally resolved stages of this period-doubling sequence provide a basis for further investigations of the predicted transition to chaos. 

\begin{figure*} [htpb]
\begin{center}
\epsfig{file=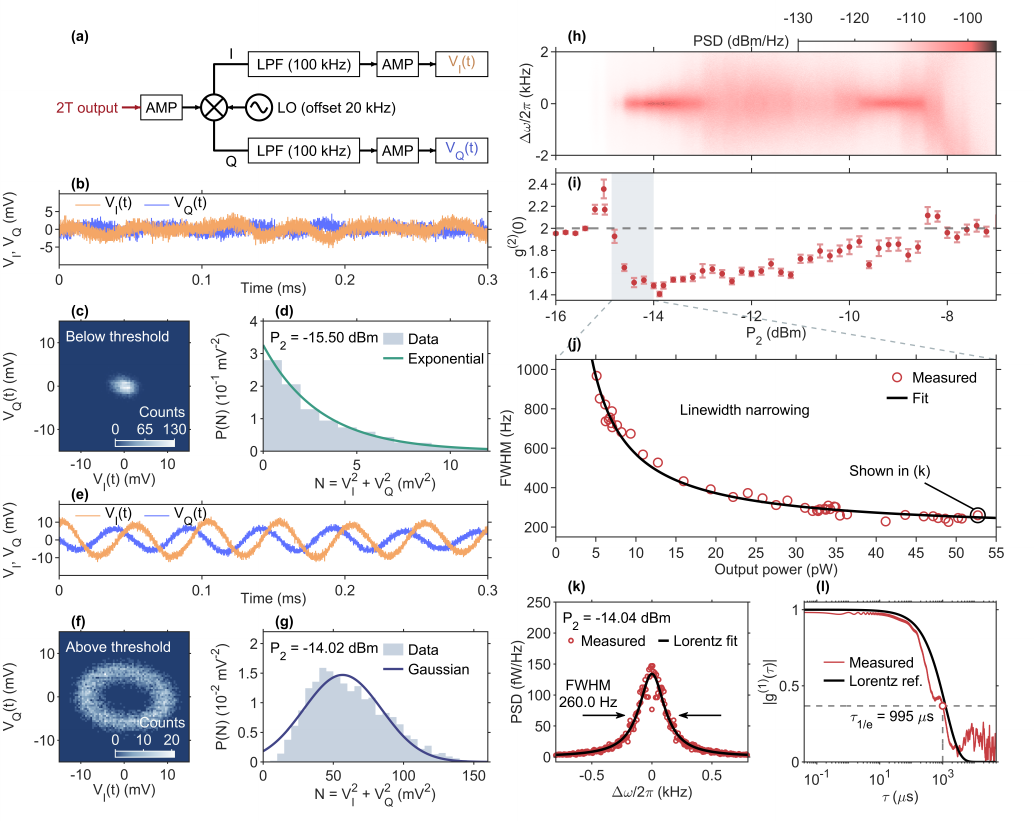,width=17.5cm}  \caption{\textbf{The centre subharmonic mode develops a narrow-line coherent self-oscillation.}
\textbf{a,} Heterodyne-detection schematic for the central microwave radiation, showing a nominal LO offset of $20~\mathrm{kHz}$, $100~\mathrm{kHz}$ low-pass filters (LPF), amplifiers (AMP) and the recorded voltages $V_I(t)$ and $V_Q(t)$.
\textbf{b,c,} Voltage traces and IQ histogram at $P_2=-15.50~\mathrm{dBm}$. \textbf{e,f,} Corresponding data at $P_2=-14.02~\mathrm{dBm}$. The two groups use $f_1=2.624~\mathrm{GHz}$, $f_2=2.626~\mathrm{GHz}$ and $P_1=-13~\mathrm{dBm}$. The displayed records span $0$--$0.3~\mathrm{ms}$, with each channel mean-subtracted; no amplitude normalization or whitening is applied.
\textbf{d,g,} Corresponding probability-density histograms of $N=V_I^2+V_Q^2$, with exponential and Gaussian fits, respectively. 
\textbf{h,i,} Centre-mode PSD and zero-delay intensity correlation $g^{(2)}(0)$ from a separate power scan at $P_1=-13~\mathrm{dBm}$. The correlation is calculated from the analytic field reconstructed from each complete $100~\mathrm{ms}$ record of the $I$ channel. Every tenth power point is displayed; error bars denote $\pm1$ circular-block-bootstrap standard error using $5~\mathrm{ms}$ blocks and $2000$ resamples. The dashed line marks $g^{(2)}(0)=2$.
\textbf{j,} Linewidth versus detected output power for $44$ consecutive settings from $P_2=-14.86$ to $-14.00~\mathrm{dBm}$, with an $a/P_{\rm out}+b$ fit. $P_{\rm out}$ is obtained by integrating the nominal-RBW-normalized measured PSD over $\pm2~\mathrm{kHz}$ without background subtraction. The circled point has the largest integrated output power among the displayed settings and identifies the spectrum in \textbf{k}.
\textbf{k,} Representative spectrum at $P_2=-14.04~\mathrm{dBm}$, measured with a nominal RBW of $5~\mathrm{Hz}$. A Lorentzian-plus-background fit over $\pm2~\mathrm{kHz}$ yields a FWHM of $260.0~\mathrm{Hz}$.
\textbf{l,} Magnitude of the first-order correlation from the analytic field reconstructed from the complete $I$-channel voltage record at $P_2=-14.04~\mathrm{dBm}$. The measured first $1/e$ crossing is $995~\mu\mathrm{s}$. The black curve is the Lorentzian reference $\exp(-\pi\Delta f|\tau|)$ inferred from \textbf{k}, corresponding to a $1/e$ time of $1.22~\mathrm{ms}$.}
\label{Maser}
\end{center}
\end{figure*}

Amplitude coherence emerges in the central subharmonic radiation as the drive power increases. We resolve its in-phase ($I$) and quadrature ($Q$) components by mixing the emitted field with a local oscillator and recording both quadratures simultaneously (see setup in Fig.~\ref{Maser}(a)). The beat drive is centred at $\bar f=2.625~\mathrm{GHz}$, with $f_{\mathrm b}=2~\mathrm{MHz}$ and $P_1=-13~\mathrm{dBm}$. At $P_2=-15.50~\mathrm{dBm}$, the voltage traces are dominated by fluctuations and the IQ distribution is concentrated near the origin (Fig.~\ref{Maser}(b,c)), consistent with a response dominated by incoherent fluctuations. At $P_2=-14.02~\mathrm{dBm}$, both quadratures exhibit sustained oscillations (Fig.~\ref{Maser}(e)), and the IQ samples are concentrated around a finite-radius ring (Fig.~\ref{Maser}(f)). This concentration at a finite amplitude reflects stabilization of the oscillation amplitude, providing a signature of laser-like amplitude coherence.

The intensity statistics resolve the transition from a fluctuating background to laser-like oscillation. From the same IQ records, we construct histograms of $N=V_I^2+V_Q^2$, where $V_I$ and $V_Q$ are the mean-subtracted quadrature voltages and $N$ is proportional to the detected microwave power. Below threshold, the intensity follows an exponential distribution (Fig.~\ref{Maser}(d)), consistent with the quasi-thermal statistics of an incoherent field. Above threshold, the distribution becomes approximately Gaussian (Fig.~\ref{Maser}(g)), reflecting fluctuations around a sustained oscillation amplitude. Together with the ring IQ distribution, this evolution supports maser-like amplitude coherence with residual intensity fluctuations. 

\begin{figure*} [htpb]
\begin{center}
\epsfig{file=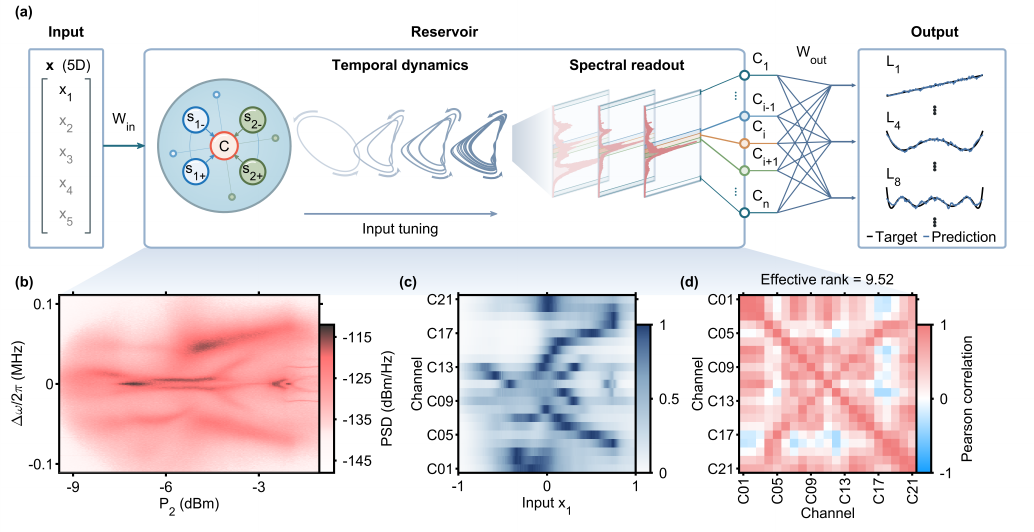,width=17.5cm}  \caption{\textbf{A magnonic time crystal as a frequency-resolved physical reservoir.} \textbf{a,} Input--reservoir--output architecture. Five accessible input dimensions correspond to pump-2 power $x_1$, pump-1--magnon detuning $x_2$, pump-1 power $x_3$, beat frequency $x_4$ and the bias-field-controlled magnon frequency $x_5$. Input tuning reshapes the interacting magnon dynamics and their temporal periodic order, illustrated schematically by phase-space trajectories, and maps the response into the emitted spectrum. Power integrated over spectral windows defines readout nodes $C_1,\ldots,C_n$. Trained output weights $W_{\rm out}$ combine their responses for function approximation. The displayed Legendre targets $L_1$, $L_4$ and $L_8$ (black) and predictions (blue) illustrate the readout following increasingly structured functions. The main text uses one-dimensional input $x_1$; two-dimensional input $(x_1,x_2)$ is presented in the Supplementary Information.
\textbf{b,} Measured radiation power spectral density versus pump-2 power $P_2$.
\textbf{c,} Normalized responses of the $21$ frequency channels versus the normalized input $x_1$.
\textbf{d,} Interchannel Pearson correlation matrix; singular-value analysis gives an effective rank of $9.52$.
}\label{nn}
\end{center}
\end{figure*}

The coherence of the central radiation is reflected in both the stability of its intensity and the persistence of its phase correlations. These complementary properties are quantified through the zero-delay intensity correlation $g^{(2)}(0)$, radiation linewidth and first-order field correlation $g^{(1)}(\tau)$. To this end, a separate sweep of $P_2$ at fixed $P_1=-13~\mathrm{dBm}$ resolves the onset, narrowing and subsequent broadening of the central line (Fig.~\ref{Maser}(h)). From each complete $100~\mathrm{ms}$ record of the in-phase ($I$) channel, we construct the analytic field $E_I(t)=v_I(t)+i\mathcal H[v_I(t)]$, where $v_I(t)$ is the mean-subtracted voltage and $\mathcal H$ denotes the Hilbert transform. With $J(t)=|E_I(t)|^2$,  we evaluate the normalized second moment of the intensity, $g^{(2)}(0)=\langle J^2\rangle/\langle J\rangle^2$ (Fig.~\ref{Maser}(i)). At low powers, the correlation is close to the thermal benchmark of two, consistent with the quasi-thermal statistics described above. Near onset, it rises to approximately $2.36$, revealing superthermal intensity fluctuations, consistent with the enhanced fluctuations that can occur near a lasing threshold\cite{G2-KAZIMIERCZUK15}. As the central line narrows, the correlation falls to approximately $1.41$, indicating increased amplitude stability. We attribute the remaining measured excess above the ideal coherent value of unity to noise added by the amplifiers in the measurement chain\cite{G2-MENZEL10}. Further increasing the drive raises the correlation again as the radiation broadens, delimiting the operating interval of enhanced amplitude coherence.

The linewidth exhibits an inverse-power narrowing trend described by $\Delta f=a/P_{\rm out}+b$, with fitted coefficients $a=3956.83~\mathrm{Hz}\cdot\mathrm{pW}$ and $b=174.51~\mathrm{Hz}$ (Fig.~\ref{Maser}(j)), consistent with the Schawlow–Townes picture\cite{LASER-SCHAWLOW58}: for comparable noise strength, a larger oscillation amplitude reduces the phase perturbation caused by a given random fluctuation, slowing phase diffusion and narrowing the radiation. This trend further supports laser coherent microwave radiation from the magnonic time crystal.

For the line-shape and temporal-coherence measurements, we select the point with the largest integrated output power among the $44$ settings in Fig.~\ref{Maser}(j), $P_2=-14.04~\mathrm{dBm}$. At this operating point, a Lorentzian fit to the central radiation yields a full width at half maximum of $\Delta f=260.0~\mathrm{Hz}$  (Fig.~\ref{Maser}(k)). The temporal coherence is quantified by $g^{(1)}(\tau)=\langle E^*(t)E(t+\tau)\rangle/\langle |E(t)|^2\rangle$, where $E(t)=v_I(t)+i\mathcal H[v_I](t)$. At the same working point, the measured $|g^{(1)}(\tau)|$ reaches $1/e$ for the first time at $\tau_{1/e}=995~\mu\mathrm{s}$, corresponding to approximately $1990$ beat periods (Fig.~\ref{Maser}(l)). This is comparable to the Lorentzian reference $1/e$ decay time, $1/(\pi\Delta f)\simeq1224~\mu\mathrm{s}$, although the measured decay is faster. The difference may reflect spectral contributions not captured by the Lorentzian fit or additional phase noise in the heterodyne readout. 

Having established temporal order and coherent central microwave radiation, we next examine how the same magnonic system can be used for information processing. The experimentally observed nonlinear spectra contain features that are not fully captured by our simplified model. This additional complexity may arise from magnon–magnon scattering channels or higher-order nonlinearities omitted from the model. These rich, input-dependent responses provide diverse nonlinear features that can be harnessed for a physical reservoir. In addition, the coherent central microwave radiation characterized above provides an externally accessible output that facilitates signal readout and offers a potential interface to other devices.

We therefore use the magnonic time crystal as a physical reservoir that maps input-dependent temporal dynamics into frequency-resolved features (Fig.~\ref{nn}(a)). The input reshapes the nonlinear dynamics of the interacting magnon modes, including their periodic response, and thereby changes the emitted spectrum. Integrating the power in fixed spectral windows produces readout nodes $C_1,\ldots,C_n$ from a single magnetic element. A trained linear output layer combines these responses to approximate a target function, $\hat y=b+\sum_{j=1}^{n}w_jF_j(\mathbf{x})$. The five accessible input dimensions are pump-2 power $x_1$, pump-1--magnon detuning $x_2$, pump-1 power $x_3$, beat frequency $x_4$ and the bias-field-controlled magnon frequency $x_5$. Here we use $x_1$ for one-dimensional input encoding; joint encoding with $(x_1,x_2)$ is presented in the Supplementary Information.

For the single-input measurement, we fix $P_1=-1~\mathrm{dBm}$ and the drive frequencies at $f_1=2.617~\mathrm{GHz}$ and $f_2=2.619~\mathrm{GHz}$, and a magnon-mode frequency $f_c =2.618~\mathrm{GHz}$. At a fixed bias field, we sweep $P_2$ in $0.05~\mathrm{dB}$ steps and select the $221$ spectra spanning $-10$ to $+1~\mathrm{dBm}$ for feature analysis; Fig.~\ref{nn}(b) displays part of this selected range. Beyond a single spectral peak associated with the $2T$ response, the output exhibits a richer spectral structure, with different frequency regions responding differently to $P_2$. In this implementation, $n=21$ fixed frequency windows define the feature vector $\mathbf{F}(\mathbf{x})=(F_1,\ldots,F_{21})$. Plotted against the linearly normalized input $x_1\in[-1,1]$, the channel responses have different shapes and turning points (Fig.~\ref{nn}(c)), providing multiple nonlinear features of the same scalar input. The channel correlation matrix and an effective rank of $9.52$ indicate that these responses are partially redundant while retaining distinct feature components (Fig.~\ref{nn}(d)).

Training only the output weights and bias, we use the spectral features to approximate Legendre functions (Fig.~\ref{nn}(a)). For the representative targets $L_1$, $L_4$ and $L_8$, the predictions follow the main variations of the target curves, including the multiple extrema of the higher-order functions.

\textit{Conclusion}.
We have realized a room-temperature magnonic time crystal with lasing coherence using a pump-induced magnon mode in YIG. Under microwave beat driving, the system exhibits period doubling over a finite power range and spectral signatures of $4T$ and $8T$ responses at stronger driving. The central radiation reaches a linewidth of $260.0~\mathrm{Hz}$, with a measured first-order correlation $1/e$ decay time of $995~\mu\mathrm{s}$. A nonlinear scattering model attributes the gain to three-magnon confluence and explains the bounded period-doubled regime through competition between scattering channels. We further use a single magnonic time crystal as a physical reservoir, mapping its input-dependent temporal dynamics into spectral features that a trained linear readout combines to approximate nonlinear functions. These advances connect microscopic magnon interactions with temporal order, lasing and nonlinear signal processing within a single magnetic element. The accessible coherent output could also mediate coupling between time-crystalline oscillators, opening opportunities to explore collective synchronization and networks for information processing.

\vskip0.25cm 

\section{Data availability}
 The data that support the findings of this study are available from corresponding authors upon reasonable request.

\section{Acknowledge}
This work has been supported by the Strategic Priority Research Program of the Chinese Academy of Sciences (Grant No. XDB0580000), STCSM (No. 23JC1404100), the National Natural Science Foundation of China (Grant Nos. 12227901, U23A6002, and 2023X0201-417-03), and start-up funding from ShanghaiTech University.

\section{Competing interests:}
The authors declare no competing interests.

\section{Correspondence:}
Correspondence and requests for materials should be addressed to  Bimu Yao, Yiming Pan and Wei Lu. 

\bibliographystyle{apsrev4-2}
\bibliography{ref}

\end{document}